\documentclass[prd,eqsecnum,showpacs,aps,nofootinbib]{revtex4}
\usepackage{latexsym}
\usepackage{graphicx} 
\def\vp{\varphi}
\def\beq{\begin{equation}}
\def\eeq{\end{equation}}
\begin{document}

\title{Unified Dark Matter in Scalar Field Cosmologies} 

\author{Daniele Bertacca}\email{daniele.bertacca@pd.infn.it}
\affiliation{Dipartimento di Fisica ``G.\ Galilei,'' Universit\`{a} di Padova, 
        and INFN, Sezione di Padova, via Marzolo 8, Padova I-35131, Italy}

\author{Sabino Matarrese}\email{sabino.matarrese@pd.infn.it}
\affiliation{Dipartimento di Fisica ``G.\ Galilei,'' Universit\`{a} di Padova, 
        and INFN, Sezione di Padova, via Marzolo 8, Padova I-35131, Italy}

\author{Massimo Pietroni}\email{massimo.pietroni@pd.infn.it}
\affiliation{INFN, Sezione di Padova, via Marzolo 8, I-35131, Italy}

\date{\today}

\begin{abstract}
Considering the general Lagrangian of \emph{k-essence} models, 
we study and classify them through variables  
connected to the fluid equation of state 
parameter $w_\kappa$. This allows to find solutions around which 
the scalar field describes a mixture of dark matter and 
cosmological constant-like dark energy, an example being the purely 
kinetic model proposed by Scherrer. Making the stronger assumption 
that the scalar field Lagrangian is exactly constant along solutions 
of the equation of motion, we find a general class of k-essence 
models whose classical trajectories directly describe a unified dark 
matter/dark energy (cosmological constant) fluid.
While the simplest case of a scalar field with canonical kinetic term 
unavoidably leads to an effective sound speed $c_s=1$, 
thereby inhibiting the growth of matter inhomogeneities,  
more general non-canonical k-essence models allow for the possibility that 
$c_s \ll 1$ whenever matter dominates. 
\end{abstract}

\maketitle

\section{Introduction}

In the current standard cosmological model, two unknown components govern 
the dynamics of the Universe: the dark matter (DM), responsible for
structure formation, and a non-zero cosmological constant $\Lambda$ 
(see, e.g. ref.~\cite{Weinberg:1988cp}), 
or a dynamical dark energy (DE) component,
that drives cosmic acceleration \cite{perlmutter,riess1,riess2}.

If the DE is given by a $\Lambda$ term, besides having 
a non-trivial fine-tuning problem to solve (unless one resorts to 
an anthropic argument), one does not know why $\Omega_{DM}$ 
and $\Omega_{\Lambda}$ are both of order unity today. 
In these years alternative routes have been followed, 
for example \emph{Quintessence} 
\cite{Wetterich88,peebles,earlyqu1,earlyqu2,earlyqu3,Carroll:1998zi,Ferreira97-1,Ferreira97-2,CLW,caldwell98,Zlatev99,Steinhardt99} 
and \emph{k-essence} \cite{Chiba:1999ka,AMS1,AMS2} (a complete list of 
dark energy models can be found in the recent 
review \cite{Copeland:2006wr}). 
The k-essence is 
characterized by a Lagrangian with non-canonical 
kinetic term and it is inspired by earlier studies of 
k-inflation \cite{kinf1,kinf2}.

Some models of k-essence have solutions which 
tend toward dynamical attractors in the cosmic 
evolution so that their late-time behavior becomes 
insensitive to initial conditions (see, e.g., 
\cite{Chiba:1999ka,Chiba:2002mw,Chimento:2003zf,Das:2006cm}). 
Other models, besides having this property allow
to avoid fine-tunning and are able to explain the cosmic 
coincidence problem \cite{AMS1,AMS2}. 
Subsequently, it was realized that the latter models  
\cite{AMS1,AMS2} have too small a basin of attraction in the radiation 
era \cite{Malquarti:2003hn} and lead to superluminal propagation of 
field fluctuations \cite{Bonvin:2006vc}. 

An important issue is whether the dark matter clustering is influenced
by the dark energy and if, when this happens, the dark energy can 
indirectly smooth the cusp profiles of dark matter at small radii. 
Another hypothesis is to consider a single fluid that behaves both as 
dark energy and dark matter. 
The latter class of models has been dubbed \emph{Unified Dark Matter} (UDM).  
Among several models of k-essence considered in the literature there 
exist two types of UDM models: \emph{generalized 
Chaplygin gas} \cite{Kamenshchik:2001cp,Bilic02,Bento:2002ps,chap_con1,chap_con2,chap_con3,chap_con4}
model and the \emph{purely kinetic model} considered by Scherrer 
\cite{Scherrer}. Alternative approaches to the unification of DM and DE 
have been proposed in Ref.~\cite{takayana}, 
in the frame of supersymmetry, and in Ref.~\cite{bono}, 
in connection with the solution of the 
strong CP problem. 

The generalized Chaplygin model can be obtained via 
a Born-Infeld-type Lagrangian. This ``fluid'' has 
the property of behaving like dark matter at high density 
and like a cosmological constant at low density. 

The kinetic model introduced by Scherrer
\cite{Scherrer} can evolve into a fluid which describes 
at the same time the dark matter and cosmological constant components.   
In this case, perturbations do not show instabilities but, at early 
times, the fluid evolves like radiation, leading to a possible 
conflict with the constraints coming from primordial 
nucleosynthesis. 
Moreover, the parameters of the model have to be fine-tuned 
in order for the model not to exhibit finite pressure effects 
in the non-linear stages of structure formation \cite{Giannakis-Hu}. 

In this paper we consider the general Lagrangian of \emph{k-essence} models 
and classify them through two variables 
connected to the fluid equation of state 
parameter $w_\kappa$. This allows to find attractor solutions around which 
the scalar field is able to describe a mixture of dark matter and 
cosmological constant-like dark energy, an example being Scherrer's 
\cite{Scherrer} purely kinetic model. 
Next, we impose that the Lagrangian of the scalar field is constant, i.e. that 
$p_\kappa=-\Lambda$, where $\Lambda$ is the cosmological constant, along 
suitable solutions of the equation of motion, and find a general class of 
k-essence models whose attractors directly describe a unified dark matter/dark 
energy fluid. 
While the simplest of such models, based on a neutral scalar field 
with canonical kinetic term, unavoidably leads to an effective speed of 
sound $c_s$ which equals the speed of light, thereby inhibiting the 
growth of matter perturbations, we find 
a more general class of non-canonical (k-essence) models which allow for 
the possibility that $c_s \ll 1$ whenever matter dominates. 

The plan of the paper is as follows. In Section 2 we introduce the general
class of k-essence models and we propose a new approach to look for 
attractor solutions. In Section 3 we apply our formalism to obtain the 
attractors for the purely kinetic case. 
In Section 4 we generalize our model giving general prescriptions 
[Eqs.~(\ref{master1}) and ~(\ref{master2})] which allow to 
obtain unified models 
where the dark matter and a cosmological constant-like dark energy 
are described by a single scalar field along its attractor solutions

Section 5 contains our main conclusions. 
The scaling solutions for a particular case of k-essence are discussed 
in Appendix A.

\section{k-essence}

Let us consider the following action 
\begin{equation}\label{eq:action}
S = S_{G} + S_{\varphi} + S_{m} = 
\int d^4 x \sqrt{-g} \left[\frac{R}{2}+\mathcal{L}(\varphi,X)\right] + S_{m}
\end{equation}
where
\begin{equation}\label{x}
X=-\frac{1}{2}\nabla_\mu \varphi \nabla^\mu \varphi \;.
\end{equation}
We use units such that $8\pi G =c^2= 1$ and signature $(-,+,+,+)$.

The energy-momentum tensor of the scalar field $\varphi$ is
\begin{equation}
  \label{energy-momentum-tensor}
  T^{\varphi}_{\mu \nu } = 
- \frac{2}{\sqrt{-g}}\frac{\delta S_{\varphi }}{\delta
    g^{\mu \nu }}=\frac{\partial \mathcal{L}
(\varphi ,X)}{\partial X}\nabla _{\mu }\varphi
  \nabla _{\nu }\varphi +\mathcal{L}(\varphi ,X)g_{\mu \nu }.
\end{equation}

If $X$ is time-like $S_{\varphi}$ describes a perfect fluid with 
$T^{\varphi}_{\mu \nu }=(\rho_{\kappa} +p_{\kappa})u_{\mu
}u_{\nu }+p_{\kappa}\,g_{\mu \nu }$, where
\begin{equation}
  \label{pressure}
  \mathcal{L}=p_{\kappa}(\varphi ,X),  
\end{equation}
is the pressure, 
\begin{equation}
  \label{energy-density}
  \rho_{\kappa} =\rho_{\kappa} 
(\varphi ,X)\equiv 2X\frac{\partial p_{\kappa}(\varphi ,X)}
  {\partial X}-p_{\kappa}(\varphi ,X)  
\end{equation}
is the energy density and the four-velocity reads 
\begin{equation}
  \label{eq:four-velocity}
  u_{\mu }= \frac{\nabla _{\mu }\varphi }{\sqrt{2X}}.
\end{equation}

Now let us assume that our scalar field defines a homogeneous background 
$X=\frac{1}{2}\dot{\varphi}^2$ (where the dot denotes differentiation w.r.t. 
the cosmic time t) 
and consider a flat Friedman-Robertson-Walker background metric.
In such a case, the equation of motion for the homogeneous mode $\varphi(t)$ 
becomes 
\begin{equation}
 \label{eq_phi}
\left(\frac{\partial p_{\kappa}}{\partial X} 
+2X\frac{\partial^2 p_{\kappa}}{\partial X^2}\right)\ddot\varphi
+\frac{\partial p_{\kappa}}{\partial X}(3H\dot\varphi)+
\frac{\partial^2 p_{\kappa}}{\partial \varphi \partial X}\dot\varphi^2
-\frac{\partial p_{\kappa}}{\partial \varphi}=0 \;. 
\end{equation}

The k-essence equation of state $w_{\kappa} 
\equiv p_{\kappa}/\rho_{\kappa}$ is just
\begin{equation}
\label{w}
w_{\kappa} = \frac{p_\kappa}{2X \frac{\partial p_\kappa}{\partial X} - 
p_\kappa} \;, 
\end{equation} 
while the effective speed of sound, which is the quantity relevant for the 
growth of perturbations, reads \cite{kinf1,kinf2} 
\begin{equation}
\label{cs}
c_s^2 \equiv  \frac{(\partial p_{\kappa} /\partial X)}
{(\partial \rho_{\kappa}/\partial X)} = 
\frac{\frac{\partial p_\kappa}{\partial X}}{\frac{\partial p_\kappa}
{\partial X}+ 2X\frac{\partial^2 p_\kappa}{\partial X^2}} \;. 
\end{equation}

If we assume that the scalar field Lagrangian depends 
separately on $X$ and $\varphi$, i.e. that it can be written in the 
form 
\begin{equation}
 \label{p_phi}
  p_{\kappa}(\varphi ,X)=f(\varphi)g(X) \; ,  
\end{equation}
then Eq.~(\ref{energy-density}) becomes 
\beq
  \label{energy-density2}
 \rho_{\kappa} = f(\varphi) \left[2\,X\,\frac{d g(X)}
  {d X}-g(X) \right]  \equiv f(\vp) \beta(X)\,. 
\eeq
Notice that the requirement of having a positive
energy density imposes a constraint on the function $g$, namely,
\begin{equation}
2X \frac{d g}{d X} > g\,,
\end{equation}
having assumed $f > 0$. 

Defining now the variables $\lambda=(1/f) df/dN$ and $\alpha= - d\ln\beta/dN$,
 where $N=\ln a$, we can express the energy density as 
\begin{equation}
\label{density_k}
 \rho_{\kappa} = \bar{K}  e^{ - \int^N dN' 
\left( \alpha(N')-\lambda(N')\right)} 
= \bar{K} e^{ - 3 \int^N dN' (w_\kappa(N') + 1)},
\end{equation}
with $\bar{K}$ an integration constant. We can also rewrite the energy 
continuity equation in the form
\begin{equation}
 \label{eq_phi2}
 \frac{d\beta}{dN} + \lambda \beta + 6 X \frac{d f}{dX} =0\,. 
\end{equation}

In terms of $\alpha$ and $w_\kappa$, or, equivalently,
of $\alpha$ and $\lambda$, the effective speed of sound, Eq.~(\ref{cs}), reads
\begin{equation}
c_s^2 =-\frac{(w_{\kappa}+1)}{2\alpha} \frac{d \ln X}{d N} 
= -\frac{\alpha - \lambda}{6\alpha} \frac{d \ln X}{d N} \; .
\end{equation}
Therefore, in purely kinetic models ($\lambda=0$) $X$ can only 
decrease in time down to its minimum value. 

The case $\alpha=0$ is analyzed in Appendix A.   

\section{Study of the attractors for purely kinetic scalar field Lagrangians}

In the $\lambda = 0$ case the Lagrangian $\mathcal{L}$ (i.e. the pressure
$p_{\kappa}$) depends 
only on $X$, that is we are recovering the equations that describe  the
purely kinetic model studied in Ref.~\cite{Scherrer} and the 
Generalized Chaplygin gas \cite{Bilic02,Bento:2002ps}.  
In this section we want to make a general study of the 
attractor solutions in this case.

For $\lambda = 0$, Eq.~(\ref{eq_phi2}) gives the following nodes,
\begin{equation}
\label{}
1) \quad \quad
\frac{d g}{d X}\biggl|_{\widehat{X}} = 0 \;, \quad \quad \quad \quad \quad 
2) \quad \quad 
X=\widehat{X} = 0 \;,  
\end{equation}
with $\widehat{X}$ a constant. 
Both cases correspond to $w_{\kappa}=-1$, as one can read from Eq.~(\ref{w}). 

The general solution of the differential equation 
(\ref{eq_phi2}) in  the  $\lambda\to 0$ limit is  \cite{Scherrer} 
\begin{equation}
\label{sol_back}
X\left(\frac{d g}{d X}\right)^{2} =  k\, e^{-6 N}
\end{equation}
with $k$ a positive constant. This solution was previously derived 
although in a different form in Ref.~\cite{chimen2}.  
As $N \rightarrow \infty$, $X$ or $d g/d X$  
(or both) must tend to zero, which shows that, depending on the specific 
form of the function $g(X)$, each particular solution will converge 
toward one of the nodes above.
 
In what follows we will provide some examples of stable node solutions of
the equation of motion, some of which have been already studied in the 
literature. The models below are classified on the basis of the stable node to 
which they asymptotically converge. 

\subsection{Case 1): Scherrer solution}

For the solution of case 1) we want to study 
the function $g$ around some $X=\widehat{X} \neq 0$. 
In this case one can approximate
$g$ as a parabola with $\frac{d g}{d X}\mid_{\widehat{X}} = 0$
\begin{equation}
\label{g}
g = g_{0}+g_{2}(X-\widehat{X})^{2}.
\end{equation}
with $g_0$ and $g_2$ suitable constants. This solution, with 
$g_0<0$ and $g_2>0$, coincides with the model studied by
Scherrer in Ref.~\cite{Scherrer}. 

If we now impose that today $X$ is close to $\widehat{X}$ so that 
$\epsilon \equiv (X-\widehat{X})/\widehat{X} = (a/a_1)^{-3} \ll 1$, 
we obtain
\begin{equation}
\label{rho_back-approx}
\rho_{\kappa} = -g_0 + 4g_2 \widehat{X}^2  \left(\frac{a} {a_1}\right)^{-3}.
\end{equation}

In order for the density to be positive at late times, we need to  
impose $g_0 < 0$. 
In this case the speed of sound (\ref{cs}) turns out to be 
\begin{equation}
c_s^2 =  \frac{(X-\widehat{X})}{(3X - \widehat{X})} = \frac{1}{2} 
\left(\frac{a} {a_1}\right)^{-3} \;, 
\end{equation}
We notice also that, for  $\left(a / a_1 \right)^{-3}\ll 1$ 
we have $c_s^2 \ll 1$.

Eq.~(\ref{rho_back-approx}) tells us that the background energy density 
can be written as $\rho_{\kappa} = \rho_\Lambda + \rho_{\rm DM}$, 
where $\rho_\Lambda$ behaves like a ``dark energy'' component 
($\rho_\Lambda = {\rm const.}$) and 
$\rho_{\rm DM}$ behaves like a ``dark matter'' component 
($\rho_{\rm DM}\propto a^{-3}$).  
Note that, from Eq.~(\ref{rho_back-approx}), $\widehat{X}$ must be 
different from zero in order for the matter term to be there. 
For this particular case the Hubble parameter $H$ is a function only 
of the UDM fluid $H^2 = \rho_\kappa/3$.

It is interesting to notice that an alternative model, proposed in 
Ref.~\cite{Novello} in the frame of extended Born-Infeld dynamics,  
actually converges to the Scherrer solution in the regime 
$(X-\widehat{X})/\widehat{X} \ll 1$.

It is immediate to verify that in the early universe case
($X \gg \widehat{X}$ i.e. $\rho_{\kappa} \gg (-g_0)$) the k-essence 
behaves like radiation. Scherrer \cite{Scherrer} imposed the constraint  
$\epsilon_0 =\epsilon(a_0) = - g_0/ ( 8 g_2 \widehat{X}^2) \ll 10^{-10}$, 
requiring that the k-essence behaves as a matter component at least 
from the epoch of matter-radiation equality. 
The stronger bound $\epsilon_0 \leq 10^{-18}$ 
is obtained by Giannakis and Hu \cite{Giannakis-Hu}, who considered the 
small-scale constraint that enough low-mass dark matter halos are produced
to reionize the universe.
One should also consider the usual constraint imposed by primordial 
nucleosynthesis on extra radiation degrees of freedom, which however leads 
to a weaker constraint. 

\subsection{Case 1): Generalized Scherrer solution}

Starting from the condition that we are near the attractor 
$X=\widehat{X}\neq 0$, we can generalize the definition of $g$, extending 
the Scherrer model in the following way
\begin{equation}
\label{p_k-n}
p_{\kappa} = g = g_0 + g_n (X - \widehat{X})^n
\end{equation}
with $n \geq 2$ and $g_0$ and $g_n$ suitable constants.\\
The density reads 
\begin{equation}
\label{rho_n}
\rho_{\kappa} = (2n-1) g_n (X-\widehat{X})^n + 2 \widehat{X} n g_n (X-\widehat{X})^{n-1} - g_0
\end{equation}
If $\epsilon^n = [(X-\widehat{X})/\widehat{X}]^n \ll 1$, 
Eq.~(\ref{sol_back}) reduces to
\begin{equation}
\label{epsilon_n-1}
X = \widehat{X} \left[1 +  \left(\frac{a} 
{a_{n-1}}\right)^{-3/(n-1)} \right]
\end{equation}
(where $a_{n-1} \ll a$) and so $\rho_{\kappa}$ becomes
\begin{equation}
\label{rho_n-epsilon}
\rho_{\kappa} \simeq  2 n \widehat{X}^n  g_n  \left(\frac{a} 
{a_{n-1}}\right)^{-3}- g_0
\end{equation}
with $(1/a_{n-1})^{-3} = [1/(n g_n)](k / \widehat{X}^{2n-1})^{1/2}$ for 
$\epsilon^n \ll 1$.

We have therefore obtained the important result that this attractor 
leads exactly to the same terms found in the 
purely kinetic model of Ref.~\cite{Scherrer}, i.e. a cosmological constant 
and a matter term. 
We can therefore extend the constraint of Ref.~\cite{Scherrer} 
to this case obtaining $(\epsilon_0)^{n-1} = - g_0 / 
(4 n \widehat{X}^n  g_n) \leq 10^{-10}$. A stronger constraint would clearly also 
apply to our model by considering the small-scale constraint 
imposed by the universe reionization, as in Ref.~\cite{Giannakis-Hu}.

If we write the general expressions for $w_{\kappa}$ and $c_s^2$ 
we have 
\begin{equation}
\label{w_n}
w_{\kappa} = - \biggl[1 + \left(\frac{g_n}{g_0}\right) (X-\widehat{X})^n \biggr]
\biggl[ 1 - 2 n \widehat{X} \left(\frac{g_n}{g_0}\right) (X-\widehat{X})^{n-1}  
- (2n-1) \left(\frac{g_n}{g_0}\right) (X-\widehat{X})^n \biggr]^{-1}
\end{equation}
\begin{equation}
\label{c2_n}
c_s^2 = \frac{(X-\widehat{X})}{2 (n-1) \widehat{X} + (2n-1) (X-\widehat{X})}.
\end{equation}

For $\epsilon \ll 1$ we obtain a result similar to that of 
Ref.~\cite{Scherrer}, namely
\begin{equation}
\label{w_n-epsilon}
w_{\kappa} \simeq -1 + 2 n \left(\frac{g_n}{ \mid g_0 \mid}\right) 
 \left(\frac{a} {a_{n-1}}\right)^{-3} \;, 
\end{equation}
\begin{equation}
\label{c2_n-epsilon}
c_s^2 \simeq \frac{1}{2(n-1)} \epsilon \;. 
\end{equation}
On the contrary, when $X \gg \widehat{X}$ we obtain 
\begin{equation}
\label{w_n-big}
w_{\kappa} \simeq c_s^2 \simeq  \frac{1}{2n-1}
\end{equation}

In this case we can impose a bound on $n$ so that at early times and/or at 
high density the k-essence evolves like dark matter. 
In other words, when $n\gg 1$, unlike the purely kinetic case of 
Ref.~\cite{Scherrer}, the model is well behaved also at high densities. 

\subsection{Case 2): Generalized Chaplygin gas}

An example of case 2) is provided by the Generalized Chaplygin (GC) 
model (see e.g. 
Refs.~\cite{Kamenshchik:2001cp,Bilic02,Bento:2002ps,chap_con1,chap_con2,chap_con3,chap_con4}),  
whose equation of state has the form 
\begin{equation} 
\label{}
p_{GC}= -\rho_* \left(\frac{\rho_{GC}}{-p_*}\right)^{\frac{1}{\gamma}} \;,
\end{equation}
where now $p_{GC}=p_{\kappa}$ and $\rho_{GC}=\rho_{\kappa}$ and 
$\rho_*$ and $p_*$ are suitable constants. 

Through the equation $\rho_{\kappa} = 2X\frac{d g(X)}{d X}-g(X)$
and the continuity equation $\frac{\rho_{GC}}{dN} + (\rho_{GC} + p_{GC}) = 0$ 
we can write $p_{GC}$ and $\rho_{GC}$ as functions of either $X$ or $a$.
When the pressure and the energy density are considered as functions 
of $X$ one gets \cite{Bento:2002ps}
\begin{equation}
\label{}
p_{GC} = -\left(\frac{-p_*}{\rho_{*}^{\gamma}}\right)^{1/(1-\gamma)} 
\left[1 - \mu X^{\frac{1-\gamma}{2}} \right]^{\frac{1}{1-\gamma}}
\end{equation}
\begin{equation}
\label{}
\rho_{GC} = \left(\frac{-p_*}{\rho_{*}^{\gamma}}\right)^{1/(1-\gamma)} 
\left[1 - \mu X^{\frac{1-\gamma}{2}} \right]^{\frac{\gamma}{1-\gamma}}
\end{equation}
with $\mu = {\rm const.}$. 

It is necessary for our scopes to consider the case 
$\gamma < 0$, so that $c_s^2 >0$.
Note that $\gamma = - 1$ corresponds to the standard ``Chaplygin gas'' model. 

Another model that falls into this class of solution is the one proposed in 
Ref.~\cite{Chimento:2003zf}, in which $g=b\sqrt{2X} - \Lambda$ (with $b$ a 
suitable constant) which 
satisfies the constraint that $p=-\Lambda$ along the attractor solution 
$X=\widehat{X}=0$. This model, however is well-known to imply a diverging speed of 
sound.

\section{Unified Dark Matter from a scalar field with 
non-canonical kinetic term} 

Starting from the barotropic equation of state $p=p(\rho)$ we can 
describe the system either through a purely kinetic k-essence 
Lagrangian (if the inverse function $\rho=\rho(p)$ exists) or through 
a Lagrangian with canonical kinetic term, as in quintessence-like models. 
The same problem has been solved in 
Ref.~\cite{Malquarti:2003nn}, although with a different 
procedure and for a different class of models. 
In the first case we have to solve the equation 
\begin{equation}
\rho(p(X))=2X \frac{d p(X)}{d X} - p(X)
\end{equation}
when $X$ is time-like. In the second case we have to solve the two 
differential equations
\begin{eqnarray}
\chi - V(\phi) & = & p(\phi,\chi) \\
\chi + V(\phi) & = & \rho(\phi,\chi)
\end{eqnarray}
where $\chi=\dot{\phi}^2/2$ is time-like. In particular if we assume that our
model describes a unified dark matter/dark energy fluid we can proceed as 
follows:
starting from $\dot{\rho}=-3H(p+\rho)=-\sqrt{3 \rho}(p+\rho)$ and  
$2 \chi =(p+\rho)=(d \phi/d \rho)^2 \dot{\rho}^2 $ we get
\begin{equation}
\label{phi_rho}
\phi = \pm \frac{1}{\sqrt{3}} \int^{\rho}_{\rho_0}
\frac{ d \rho' / \sqrt{\rho'}}{(p(\rho')+\rho')^{1/2}} \; ,
\end{equation}
up to an additive constant which can be dropped without any 
loss of generality. 
Inverting the Eq.~(\ref{phi_rho}) i.e. writing $\rho=\rho(\phi)$ 
we are able to get $V(\phi)=[\rho(\phi)-p(\rho(\phi))]/2$. 
If one requires the exact condition that our unified DM fluid has a 
constant pressure term $p=-\Lambda$ and looks for a scalar field model 
with canonical kinetic term, one arrives at an exact solution
with potential $V(\phi)=(\Lambda/2)[\cosh^2({\sqrt 3}\phi/2) + 1 ]$ (see 
also Ref.~\cite{stability1,stability2}). 
Using standard criteria (e.g. Ref.~\cite{Copeland:2006wr})  
it is immediate to verify that the above trajectory corresponds to 
a stable node even in the presence of an extra-fluid (e.g. radiation) with
equation of state $w_{\rm fluid}\equiv p_{\rm fluid}/\rho_{\rm fluid}>0$, 
where $p_{\rm fluid}$ and $\rho_{\rm fluid}$ are the fluid pressure 
and energy density, respectively. 
Along the above attractor trajectory our scalar field behaves 
precisely like a mixture of pressure-less matter and cosmological constant. 
Using the expressions for the energy density and the pressure we 
immediately find, for the matter energy density 
\begin{equation}
\rho_m = \rho - \Lambda = \Lambda \sinh^2 \left(\frac{\sqrt 3}{2} \phi \right) 
\propto a^{-3} \;.
\end{equation} 
A closely related solution was found by Salopek \& Stewart \cite{salste}, 
using the Hamiltonian formalism. 
Like any scalar field with canonical kinetic term \cite{Bludman}, 
such a UDM model however predicts $c_s^2=1$, as it is clear from 
Eq.~(\ref{cs}), which inhibits the growth of matter inhomogeneities. 
Such a ``quartessence'' model therefore behaves exactly 
like a mixture of dark matter and dark energy along the attractor solution, 
whose matter sector, however is unable to cluster on sub-horizon scales 
(at least as long as linear perturbations are considered). 

We can then summarize our findings so far by stating 
that purely kinetic k-essence
cannot produce a model which {\it exactly} describes a unified fluid of 
dark matter
and cosmological constant, while scalar field models with canonical kinetic 
term, while containing such an exact description, unavoidably lead to 
$c_s^2=1$, in conflict with cosmological structure formation. 
In order to find an exact UDM model with acceptable speed of sound we 
consider more general scalar field Lagrangians. 

\subsection{Lagrangians of the type 
${\mathcal L}(\varphi,X) = g(X) - V(\varphi)$}

Let us consider Lagrangians with non-canonical kinetic term and 
a potential term, in the form 
\begin{equation}
\label{newlagr}
{\mathcal L}(\varphi,X) = g(X) - V(\varphi) \;. 
\end{equation}
The energy density then reads 
\begin{equation}
  \label{newenergy-density}
  \rho = 2X\frac{d g(X)}
  {d X}-g(X) + V(\varphi) \;,  
\end{equation} 
while the speed of sound keeps the form of Eq.~(\ref{cs}). 
The equation of motion for the homogeneous mode reads 
\begin{equation}
\label{new_eqmot}
\left(\frac{d g}{d X}+
2X\frac{d^2 g}{d X^2} \right)\frac{dX} {dN} 
+ 3 \left(2 X \frac{d g}{d X}\right) = -  \frac{dV}{dN} \;. 
\end{equation}

One immediately finds 
\begin{equation}
p + \rho  = 2X\frac{d g(X)}{d X} \equiv 2{\cal F}(X) \;. 
\end{equation} 

We can rewrite the equation of motion, Eq.~(\ref{new_eqmot}), 
in the form 
\begin{equation}
\left[ 2X \frac{d {\cal F}}{d X} - {\cal F} \right]
\frac{dX} {dN} + X \left( 6{\cal F} + \frac{dV}{dN} \right) = 0 \;. 
\end{equation}
It is easy to see that this equation admits 2 nodes, namely: 1) 
$d g/d X|_{\widehat{X}} =0$ and 2) $\widehat{X}=0$. In all cases, 
for $N \to \infty$, the potential 
$V$ should tend to a constant, while the kinetic term can be
written around the attractor in the form 
\begin{equation}
\label{xn}
g(X) = M^4 \left(\frac{X - \widehat{X}}{M^4}\right)^n \quad \quad \quad n \geq 2 \;, 
\end{equation}
with $M$ a suitable mass-scale and the constant $\widehat{X}$   
can be either zero or non-zero. The trivial case $g(X)=X$ obviously 
reduces to the one of Section 4. 
 
Following the same procedure adopted in the previous section 
we impose the constraint $p=-\Lambda$, which yields the general solution
$\rho_m=2{\cal F}(X)$. 

This allows to define $\varphi=\varphi(\rho_m)$ as a solution of the 
differential equation 
\begin{equation}
\label{master1}
\rho_m = 2{\cal F}\left[\frac{3}{2}\left(\rho_m +\Lambda\right) \rho_m^2 
\left(\frac{d \varphi}{d \rho_m}\right)^2 \right] \;.  
\end{equation}

As found in the case of k-essence, the most interesting behavior 
corresponds to the limit of large $n$ and $\widehat{X}=0$ in Eq.~(\ref{xn}), 
for which we obtain 
\begin{equation}
\rho_m \approx \Lambda \sinh^{-2}\left[\left(\frac{3\Lambda}{8M^4} \right)
^{1/2} \varphi\right] \;,
\end{equation}
leading to 
$V(\varphi) \approx \rho_m/2n - \Lambda$, and $c_s^2 = 1/(2n-1) \approx 0$. 
The Lagrangian of this model is similar to that analyzed in 
Ref.~\cite{DiezTejedor-Feinstein}.\\

\subsection{Lagrangians of the type 
${\mathcal L}(\varphi,X) = f(\varphi) g(X)$}

Let us now consider Lagrangians with a non-canonical kinetic term 
of the form of Eq.~(\ref{p_phi}), namely 
${\mathcal L}(\varphi,X) = f(\varphi)g(X)$. 

Imposing the constraint $p=-\Lambda$, we obtain $f(\varphi)=-\Lambda /g(X)$,
which inserted in the equation of motion yields the general solution
\begin{equation}
\label{master2}
X \frac{d \ln g}{dX} = - \frac{\rho_m}{2\Lambda} \;. 
\end{equation}

The latter equation, together with Eq.~(\ref{master1}) define our general
prescription to get UDM models describing both DM and cosmological 
constant-like DE. 

As an example of the general law in Eq.~(\ref{master2})
let us consider an explicit solution. Assuming that  
the kinetic term is of Born-Infeld type, as in 
Refs.~\cite{prc,abramo1,abramo2,stability1,stability2},  
\begin{equation}
g(X)=-\sqrt{1-2X/M^4} \;, 
\end{equation}
with $M$ a suitable mass-scale, which implies $\rho = 
f(\varphi)/\sqrt{1-2X/M^4}$, we get
\begin{equation}
\label{udmsol_X}
X(a) =\frac{M^4}{2}\frac{\bar{k} a^{-3}}{1 + \bar{k} a^{-3}} \;,
\end{equation}
where $\bar{k}=\rho_m(a_*) a_*^3/ \Lambda $ and $a_*$ is the scale-factor at 
a generic time $t_*$. 
In order to obtain an expression for $\varphi(a)$, we impose that the 
Universe is dominated by our UDM fluid, i.e. $H^2=\rho/3$. This gives 
\begin{equation}
\label{udmsol}
\varphi(a) =\frac{2 M^2}{\sqrt{3\Lambda}} 
\left\{ \arctan \left[\left(\bar{k} a^{-3}\right)^{-1/2}\right] 
- \frac{\pi}{2} \right\} \;, 
\end{equation}
which, replaced in our initial ansatz $p=-\Lambda$ allows to obtain the 
expression (see also Ref.~\cite{stability1,stability2})
\begin{equation}
f(\varphi)=\frac{\Lambda}{\left\vert\cos 
\left[ \left(\frac{3\Lambda}{4M^4}\right)^{1/2}\varphi \right] 
\right\vert} \;.
\end{equation}

If we expand $f(\varphi)$ around $\varphi= 0$,  
and $X/M^4 \ll 1$ we get the approximate Lagrangian
\begin{equation}
{\mathcal L}  \approx \frac{\Lambda}{2M^4}\dot{\varphi}^2 - 
\Lambda \left[1 + \frac{3\Lambda}{8M^4}\varphi^2\right] \;.
\end{equation}
Note that our Lagrangian depends only on the combination $\varphi/M^2$, 
so that one is free to reabsorb a change of the mass-scale in the 
definition of the filed variable. 
Without any loss of generality we can then set $M=\Lambda^{1/4}$,  
so that the kinetic term takes the canonical form in the limit $X\ll 1$. 
We can then rewrite our Lagrangian as 
\begin{equation}
\label{udmbest}
{\mathcal L} = - \Lambda \frac{\sqrt{1-2X/\Lambda} }
{\left\vert\cos \left(\frac{\sqrt{3}}{2} \varphi \right)\right\vert } \;. 
\end{equation}

This model implies that for values of 
$\sqrt{3}\varphi \approx -\pi$ and $2X/\Lambda \approx 1$, 
\begin{equation}
\cos \left(\frac{\sqrt{3}}{2} \varphi \right)  \propto  a^{3/2}\;,
\quad \quad \quad  \sqrt{1-2X/\Lambda}  \propto  a^{-3/2} \;,
\end{equation}
the scalar field mimics a dark matter fluid. In this regime the effective 
speed of sound is $c_s^2=1-2X/\Lambda \approx 0$, as desired. 

To understand whether our scalar field model gives rise to a 
cosmologically viable UDM solution, we need to check if in a Universe 
filled with a scalar field with Lagrangian (\ref{udmbest}), plus 
a background fluid of e.g. radiation, the system displays the desired
solution where the scalar field mimics both the DM and DE components. 
Notice that the model does not contain any free parameter to specify 
the present content of the Universe. 
This implies that the relative amounts of DM and DE that characterize 
the present universe are fully determined by the value of 
$\varphi_0\equiv \varphi(t_0)$. In other words, to reproduce the present
universe, one has to tune the value of $f(\varphi)$ in the early Universe. 
However, a numerical analysis shows that, once the initial value of 
$\varphi$ is fixed, there is still a large basin of attraction in terms of the 
initial value of $d\varphi / dt$, which can take any value such that 
$2X/\Lambda \ll 1$. 

\begin{figure}[t]
\includegraphics[width = 4in , keepaspectratio=true]{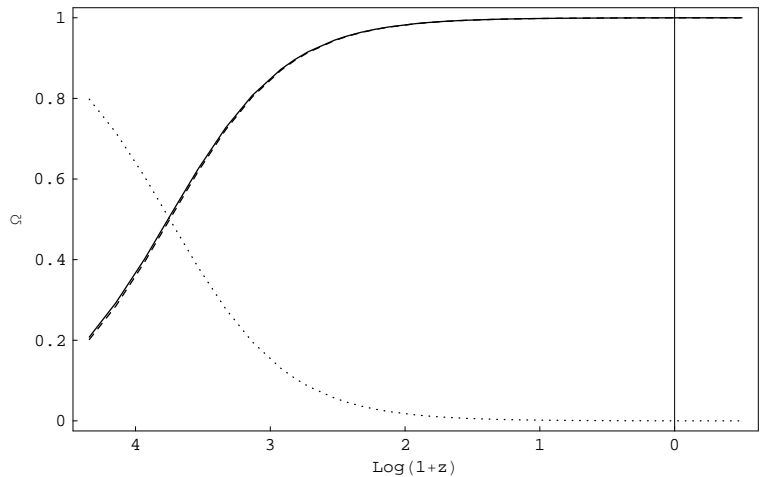}
\caption{Evolution of the scalar field density parameter vs. redshift. 
The continuous line shows the UDM density parameter; the dashed line is the 
density parameter of the DM + DE components in a standard $\Lambda$CDM model; 
the dotted line is the radiation density parameter.}
\label{fig1}
\end{figure}
\begin{figure}[t]
\includegraphics[width = 4in,keepaspectratio=true]{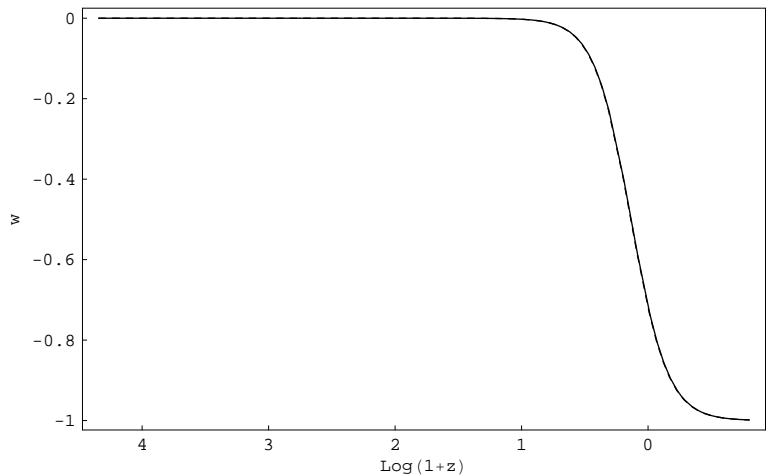}
\caption{The redshift evolution of the scalar field equation of state 
parameter $w_{\rm UDM}$ (continuous line) is compared with that of the 
sum of the DM + DE components in a standard $\Lambda$CDM model 
(dashed line).}
\label{fig2}
\end{figure}

The results of a numerical integration of our system including scalar 
field and radiation are shown in Figures 1 - 3. Figure 1 shows the 
density parameter, $\Omega_{\rm UDM}$ as a function of redshift, having 
chosen the initial value of $\varphi$ so that today the 
scalar field reproduces the observed values $\Omega_{\rm DM} \approx 0.268$,
$\Omega_{\rm DE} \approx 0.732$ \cite{Spergel}. 
Notice that the time evolution of the scalar field energy density 
is practically indistinguishable from that of a standard DM plus Lambda 
($\Lambda$CDM) model with the same relative abundances today.    
Figure 2 shows the evolution equation of state parameter $w_{\rm UDM}$; 
once again the behavior of our model is almost identical to that of 
a standard $\Lambda$CDM model for $1 + z < 10^4$. Notice that, since 
$c_s^2=-w_{\rm UDM}$, the effective speed of sound of our model
is close to zero, as long as matter dominates, as required. 
In Figure 3 we finally show the redshift evolution of the scalar field 
variables $X={\dot \varphi}^2/2$ and $\varphi$: one can easily check that
the evolution of both quantities is accurately described by the analytical 
solutions above, Eqs.~(\ref{udmsol_X}) and (\ref{udmsol}), respectively (the 
latter being obviously valid only after the epoch of matter-radiation 
equality). 

\begin{figure}
\includegraphics[width = 4in,keepaspectratio=true]{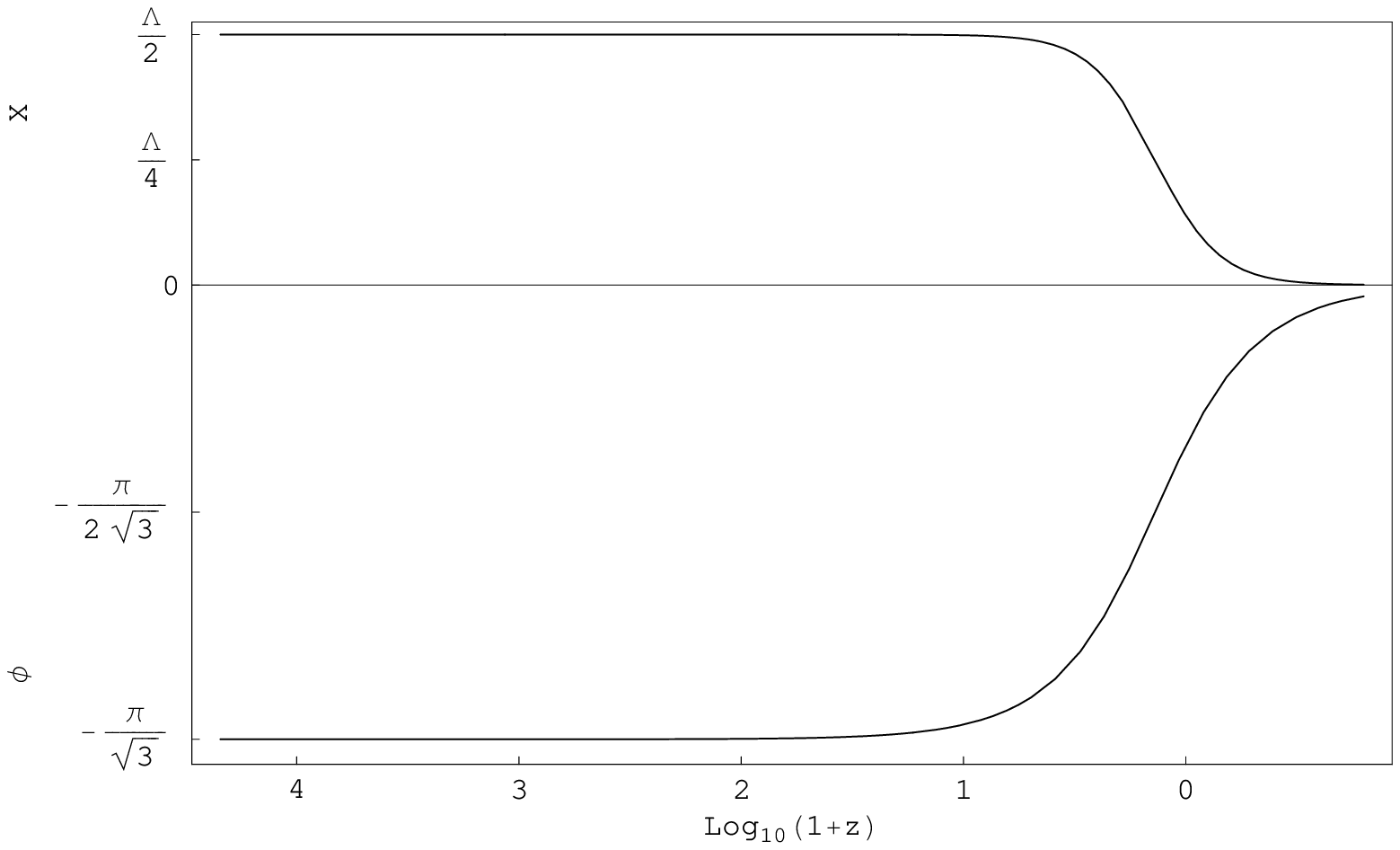}
\caption{Redshift evolution of the scalar field 
of the scalar field variables $X={\dot \varphi}^2/2$ (top) and $\varphi$
(bottom).} 
\label{fig3}
\end{figure}

\section{Conclusions}

In this paper we have investigated the possibility that the dynamics of 
a single scalar field can account for a unified description of the 
dark matter and dark energy sectors. In particular, 
we have studied the case of purely kinetic k-essence, 
showing that these models have only one late-time attractor with 
equation of state $w_{\kappa} = -1$ (cosmological constant). 
Studying all possible solutions near the attractor we have 
found a generalization of the Scherrer model \cite{Scherrer}, which 
describes a unified dark matter fluid. 

Generalizing our analysis to the case where the Lagrangian is not 
purely kinetic, we have given general prescriptions [Eqs.~(\ref{master1})
and ~(\ref{master2})] to obtain unified models where the dark matter 
and a cosmological constant-like dark energy are described 
by a single scalar field along its attractor solutions. Moreover, we have 
given explicit examples for which the effective speed of sound is 
small enough whenever
matter dominates, thus allowing for the onset of gravitational instability. 
Studying the detailed consequences of such unified dark matter 
model for cosmic microwave background anisotropies and for the formation 
of the large-scale structure of the Universe will be the subject of a 
subsequent analysis. 

\section*{Acknowledgments}
We thank Nicola Bartolo and Fabio Finelli for useful discussions. 

\vskip 1cm
\appendix
\setcounter{equation}{0}
\def\theequation{A.\arabic{equation}}
\vskip 0.2cm
\section{Study of the solutions for $\alpha = 0$} 

If $X$ is constant ($dX/dN$ = 0) then, from Eq.~(\ref{eq_phi2}),  
$\alpha = 0$. In this situation we have that 
\begin{equation}
\label{eq_f_when_alpha_is_zero}
\frac{1}{f}\frac{df}{dN}=\frac{1}{\dot{N}}\frac{1}{f}\frac{df}{dt} = 
\lambda= -3(w_{\kappa}+1)
\end{equation}
is constant because $w_{\kappa}$ is function only of $X$. 

Now if we consider the case in which the universe is dominated by a
fluid with constant equation of state $w_B$ then $H=\dot{N} \sim 
2/[3(w_B+1)t]$ and Eq.~(\ref{eq_f_when_alpha_is_zero}) becomes
$ d \ln f/d\ln t \sim -2 (w_{\kappa}+1)/(w_{B}+1)$.

Therefore we get $f \sim t^{-2 \frac{w_{\kappa}+1}{w_{B}+1}} \sim 
\varphi^{-2 \frac{w_{\kappa}+1}{w_{B}+1}}$ 
(because if $\dot{\varphi}$ is constant then $\varphi\sim \sqrt{2X} t$). 
In other words, we have recovered, although in a more general way, the 
result of Ref.~\cite{Chiba:1999ka}. 
These models have been dubbed ``scaling k-essence '' 
(see also Ref.~\cite{tsuji}).  
In such a case, the equation of state $w_{\kappa}$ can be written as 
$w_{\kappa}=\beta (w_{B}+1)/2 -1$, 
where $f=\varphi^{- \beta}$. If $w_B = w_{\kappa}$ we have only 
the k-essence as background and we get $\beta = 2$. 

Using the latter approach it is simpler to see that when 
$\alpha \rightarrow 0$ all the viable solutions 
\footnote{If $\alpha \rightarrow c > 0$ 
for $N \rightarrow +\infty$ we 
have $2 X \frac{d g}{d X} -g \rightarrow 0$ and we get three 
possible solutions: 
i) if $g \rightarrow {\rm const.}$ 
we have $w_{\kappa} \rightarrow \infty$, which 
is not acceptable; 
ii) if $g \rightarrow 0$ and $w_{\kappa} \rightarrow {\rm const.} \neq 0$ 
we get $w_{\kappa} \rightarrow c_s^2 > 0$ \cite{Das:2006cm}; 
this solution cannot be used to describe an accelerated universe; 
iii) if $g \rightarrow 0$ and $w_{\kappa} \rightarrow 0$ 
we get $0 \leq c_s^2 \leq 1$; obviously, also in this case the universe 
cannot accelerate.} converge 
to the scaling solution. In fact, if a priori $\alpha \neq 0$ and 
$f = \varphi^{-\beta}$ we have that
\begin{equation}
\label{alpha_Wb_constant}
\alpha = 3(w_{\kappa}+1) - 
\frac{3}{2}\beta (w_{B}+1)\frac{\sqrt{2X} t}{\varphi}.
\end{equation}

Starting from Eq.~(\ref{eq_phi2}), we note that when $\alpha \rightarrow 0$ 
then $2 X \frac{d g}{d X} -g \rightarrow {\rm const.} \neq 0$. 
It is then easy to see that $X$ must be a constant 
\footnote{When $2 X \frac{d g}{d X} -g = 
{\rm const.}$ we have two possibilities: 
either $X$ is constant or we need resolve 
directly this differential equation. In this last case we obtain 
$g = b\sqrt{2X} + {\rm const.}$, which implies constant $X$ 
but diverging speed of sound $c_s^2$ \cite{Chimento:2003zf}.}. 
Therefore, provided that $g \neq  b\sqrt{2X} + {\rm const.}$ 
the solution of the equation of motion converges to the scaling solution.



\end{document}